\documentclass[twocolumn,showpacs,preprintnumbers,amsmath,amssymb]{revtex4}

\begin{document}
\title{Oxygen Phonon Branches in YBa$_2$Cu$_3$O$_7$}
\author{L. Pintschovius, D. Reznik*, W. Reichardt}
\affiliation{Forschungszentrum Karlsruhe, Institut f\"ur
Festk\"orperphysik, P.O.B. 3640, D-76021 Karlsruhe, Germany}
\author{Y. Endoh, H. Hiraka}
\affiliation{Institute for Material Research, Tohoku University,
Katahira, Aoba-ku, Sendai, 980-8577, Japan}
\author{J.M. Tranquada}
\affiliation{Physics Department, Brookhaven National Laboratory, Upton,
NY 11973, USA}
\author{H. Uchiyama, T. Masui, S. Tajima}
\affiliation{Superconductivcity Research Laboratory, ISTEC, Shinonome,
Koutu-ku, Tokyo, 135-0062, Japan}
\date{\today}
\begin{abstract}
We report results of inelastic neutron scattering measurements of
phonon dispersions in optimally doped YBa$_2$Cu$_3$O$_{6.95}$ and
compare them with model calculations. The focus is on the in-plane
oxygen bond-stretching phonon branches. The study of these modes
is complicated by anticrossings with c-axis-polarized branches;
such effects are interpreted through lattice-dynamical shell-model
calculations. The in-plane anisotropy of the bond-stretching
phonons was firmly ascertained from measurements on a detwinned
sample. Studying not only the in-plane modes involving in-phase
motion for the two Cu-O layers within a unit cell but also those
with opposite-phase motion was of great help for establishing a
clear experimental picture. The measurements confirm that the
in-plane oxygen bond-stretching phonon branches disperse steeply
downwards from the zone center in both the a and the b directions
indicating a strong electron-phonon coupling. For the
b-axis-polarized bond-stretching phonons, there is an additional
feature of considerable interest: a sharp local frequency minimum
was found to develop on cooling  from room temperature to T = 10 K
at wave vector {\bf q}$\approx$0.27 r.l.u..
\end{abstract}

\pacs{74.25.Kc,63.20.Kr,74.72.Bk}
\maketitle
\newpage
\section{Introduction}
Numerous inelastic neutron scattering
investigations\cite{Pini1998,Pini1999,McQueeny} on high-T$_c$
superconductors and their insulating parent compounds have shown
that the high-energy Cu-O bond-stretching modes soften
considerably on doping. It was advocated from the very
beginning\cite{Renker} that this phonon softening indicates a
strong electron-phonon coupling, but it attracted nevertheless
only limited attention because phonons were widely considered to
be irrelevant for the mechanism of high-T$_c$ superconductivity.
Recently, renewed interest in the phonons came from observations
made by angular resolved photoemission spectroscopy of an abrupt
change of electron velocity at 50-80 meV which was interpreted as
evidence for an ubiquitous strong electron-phonon coupling in
high-T$_c$ superconductors\cite{Lanzara,Shen}. Further interest
was generated by theories\cite{Zaanen,Machida,Emery} predicting an
inhomogeneous charge distribution in the Cu-O planes, in
particular in the form of stripes. In superconducting samples,
charge stripes are assumed to be dynamic in nature which makes it
very difficult to detect them. They might reveal themselves by
coupling to the phonons\cite{Neto,Park,Kaneshita} thus making
inelastic neutron scattering an appealing technique. Naively, one
might expect precursor phenomena to charge-density-wave (CDW)
formation similar to what has been observed in one-dimensional
conductors such as KCP some time ago\cite{Renker1973}. There, a
dip develops in the phonon dispersion of the longitudinal acoustic
branch in a narrow range of momentum transfer about a wavevector
related to the spanning vector 2k$_F$ of the Fermi surface.

In the cuprates, the most pronounced effects are expected for the
Cu-O bond stretching vibrations because the Cu-O bond length will
be significantly modulated by the extra charge residing on the
stripes. Early investigations\cite{Reichardt1996} of the phonons
in YBa$_2$Cu$_3$O$_{7-x}$ have indeed shown that the O
bond-stretching vibrations within the Cu-O planes behave in a
normal way in insulating samples but in an apparently anomalous
way in optimally doped ones: the frequency of these phonons
decreases abruptly when going from the zone center along the [100]
(={\bf a*}) direction or the [010](={\bf b*}) direction about half
way to the zone boundary. In addition, the phonon peaks become
very poorly defined in this q-range. Unfortunately, twinning of
the samples did not allow one to decide whether the phonons in
question really acquire a very large intrinsic linewidth or
whether the observed peaks are so broad simply because the
superimposed contributions from the {\bf a*} and the {\bf b*}
directions differ in energy. A recent inelastic neutron scattering
investigation \cite{Chung} using pulsed neutrons came to the
conclusion that the Cu-O bond-stretching branch along the {\bf b*}
direction has a steep, but continuous downward dispersion whereas
the branch along {\bf a*} splits into two in the middle of the
zone. The latter finding was tentatively interpreted as a
signature of charge stripes. We will discuss this interpretation
in the light of our own results.

The present paper describes results of inelastic neutron
scattering measurements made by the triple-axis technique on
detwinned as well as on twinned samples of optimally doped
YBa$_2$Cu$_3$O$_{6.95}$ aiming at elucidating the anomalous
behavior of the bond-stretching modes. The measurements on the
detwinned samples were key to achieving a better understanding of
the a-b anisotropy and of the seemingly anomalous lineshapes. We
show that very complex intensity distributions can be explained as
resulting from an interaction of phonon branches having the same
symmetry but a different polarization\cite{brief}. The line
broadenings due to anti-crossings partly mask a more important
reason for anomalous lineshapes, i.e. a sharp drop of phonon
frequencies within a narrow range of wave vectors along {\bf b*}
observed at low temperatures. As has been pointed out in a
separate publication\cite{Pinicondmat} this peculiar behavior is
observed only at low temperatures and strongly suggests dynamic
charge stripe formation with a period of about four lattice
constants.

The rest of the paper is organized as follows. Experimental
details are given in the following section. The model used to
simulate the phonon dispersions and scattering intensities is
described in Sec. III. These calculations are essential for making
sense of the measurements, which are presented in Sec. IV. A
discussion of the results and conclusions are given in Sec.V and
VI, respectively. Appendix A provides a brief description of
anticrossing behavior between neighboring phonon branches. The
special force constants, beyond the shell model, required to
simulate the observations are described in Appendix B.

\section{Experimental}
The first series of measurements was performed on a detwinned
sample. It consisted of 26 detwinned single crystals co-aligned
with an effective mosaic spread of $\sim$3 degrees. The total
volume of the sample was $\sim$1 cm$^{3}$. Unfortunately, the
crystals cleaved into many small pieces after the first cooling
cycle, and therefore additional measurements had to be performed
on twinned samples. The next round of measurements was carried out
on a composite sample weighing a total of 31 g which was used
previously for a part of the pulsed neutron measurements of Chung
et al\cite{Chung}. However, this sample was not very well suited
for triple-axis measurements because the lateral dimensions were
too large compared to the size of the neutron beam. Moreover,
high-precision measurements of the c-axis lattice constant of the
individual crystals revealed that some of them were underdoped.
Based on these findings, another composite sample was assembled
out of 3 optimally doped (YBa$_2$Cu$_3$O$_{6.95}$, T$_c$ = 93 K)
single crystals of combined volume of $\sim$1.5cm$^{3}$ co-aligned
with a mosaic spread of 2.2 degrees. All the temperature dependent
studies were done on this sample.

The experiments were carried out on the triple-axis spectrometer
1T located at the Orph$\acute{e}$e reactor using doubly focusing
monochromator (Cu111 or Cu220) and analyzer (PG002) crystals.
Specifically, Cu111 was used as monochromator for the measurements
on the detwinned sample to maximize the intensity whereas Cu220
was used for the measurements on the twinned samples to achieve
high resolution. The energy resolution at energy transfer of
E$\approx$70 meV was $\Delta$E=4.6 (2.9) meV (FWHM) for the Cu111
(Cu220) monochromator and a final energy E$_F$=14.7 meV. The
measurements were carried out in different scattering planes in
order to elucidate the eigenvectors of the atomic vibrations in
question. Most measurements were performed at T=12 K but selected
phonons were also studied at higher temperatures up to T=300 K.

\section{Phonon Model Calculations}
In measurements on compounds with many atoms in the unit cell, an
assignment of phonon peaks to particular modes has to be based on
model calculations. For this purpose, extensive model calculations
were carried out prior to and in parallel with the experiments.
The model used as a starting point was the common interaction
potential model reported in Ref. 19, which is quite successful in
describing the phonon dispersion curves of a number of cuprates.
In the framework of this model, YBa$_2$Cu$_3$O$_7$ is considered
as an ionic compound, and the interatomic interactions are modeled
as a sum of Coulomb forces and short range repulsive forces. In
addition, the polarizability of the atoms is taken into account
using the shell model formalism. For a metallic compound like
YBa$_2$Cu$_3$O$_7$ a term accounting for screening by free
carriers is added. This model gives a decent description of most
of the available data shown in Ref. 1. We further improved upon
the agreement between model and experiment by tuning the
parameters; the refined parameters are listed in Table I. However,
it was already clear before the present study that extra ad hoc
interactions are required to adequately describe the oxygen
bond-stretching modes.

Experiments\cite{Pini1998,Reichardt1996} have shown that the
bond-stretching phonon frequencies are strongly renormalized. This
effect is much stronger in the [100] and in the [010] directions
than in the [110] direction.It leads to a downward dispersion from
zone center to zone boundary\cite{Reichardt1996} which cannot be
reproduced by simply tuning the parameters of the common
interaction potential model\cite{Chaplot}. Adding to the model a
negative breathing deformability lowers the calculated energy of
the planar breathing mode, which is the endpoint of the
bond-stretching phonon branch in the [110] direction. At the same
time, this term lowers the energy of the corresponding branches in
the [100] and the [010] directions, but by only half the amount of
that in the [110] direction. Therefore, additional terms were
added lowering specifically the energies of the ''half-breathing''
modes at the zone boundary of the [100] and - with a different
parameter - of the [010] direction. The extra terms are described
in detail in Appendix B.

To proceed further with this discussion, it is necessary to
illustrate the situation with actual data. Figure 1 compares
calculated dispersions with measured O bond-stretching modes
dispersing along [010] and involving opposite-phase motion for the
two Cu-O layers within a unit cell ($\Delta_4$' symmetry). The fit
to the 200-K data with the additional terms discussed above is
indicated by the magenta curve. This fit matches the data well at
the zone center and boundary, but falls significantly above the
data in the middle of the zone. To remedy this deficiency, a
further term was included that produces a maximum effect at the
halfway point but zero effect both at the zone center and at the
zone boundary. The resulting fit is indicated by the red line in
Fig. 1. An important result of the present investigation is the
observation of an anomalous softening in the [010] direction
within a rather narrow range of wave vectors on cooling from room
temperature to low temperatures. In order to fit the 10 K data, an
additional term producing a sharper effect halfway to the zone
boundary had to be included with the resulting fit indicated by
the blue line in Fig. 1. We believe that all of the extra terms we
have added reflect the electron-phonon coupling resulting from
doping holes into the CuO$_2$ planes.

Before continuing on, it may be useful to make contact with recent
theoretical work. Bohnen, Heid, and Krauss \cite{Bohnen} have
calculated the phonon dispersions for YBa$_2$Cu$_3$O$_7$ using a
method based on evaluating the band structure in the local-density
approximation (LDA). Their results successfully match the
experimental $\Delta_4$' branch at 200 K (see Ref.
\cite{Pinicondmat}) though they do not predict the temperature
induced softening halfway to the zone boundary. Further, their
calculations reproduce the corresponding $\Delta_4$ branch along
[100] very well (Fig. 8). The LDA results provide some
justification for the ad hoc interactions that have been discussed
above and point at the unconventional nature of the temperature
dependence observed in the [010] direction.

We note again that Fig. 1, which illustrates the effects of the
special terms, was prepared for branches of $\Delta_4$-symmetry.
Phonons of $\Delta_4$-symmetry are distinguished from those of
$\Delta_1$-symmetry in that the atoms in the Cu-O bi-layer move in
opposite phase instead of in-phase. For modes of
$\Delta_1$-symmetry, the downward dispersion of the
bond-stretching mode leads to anti-crossings with branches of the
same symmetry which originally have a different polarization but
that hybridize with the bond-stretching phonons where they are
close in energy (readers who are not familiar with such effects
are referred to APPENDIX A). The anti-crossings are naturally
explained by the model, and the model calculations for $\Delta_1$
frequencies and intensities agree well with the experimental
measurements, as will be shown in the next section. It happens
that the branches of $\Delta_4$-symmetry are much less affected by
anti-crossing effects.

\section{Results and Analysis}
\subsection{Low temperature measurements}
Experimental spectra obtained on the detwinned sample are plotted
in Figs. 2,3. The two series of scans were conducted along lines
from the (4,1,0) reciprocal lattice point to (3.5,1,0) and from
(1,4,0) to (1,3.5,0), respectively. These lines were chosen
because (i) the inelastic structure factors of the bond-stretching
modes are quite favorable here and (ii) these choices allowed us
to exploit focusing effects. The figures further show simulated
spectra calculated with the help of the lattice dynamical model
described in the previous section. We note that the full
resolution in q and w was taken into account for the simulations.
There is obviously very good agreement between the experimental
and the calculated spectra. Another way to visualize this
agreement is to represent both types of spectra as (color coded)
contour plots (Figs. 4,5). The plots computed from the simulated
spectra also show the phonon dispersion curves calculated from the
same model for the high symmetry directions. All the depicted
branches are of the same symmetry, i.e. $\Delta_1$ (Figs. 2,4) and
$\Delta_1$'(Figs. 3,5), respectively (in order to distinguish
phonon branches along {\bf a*} from those along {\bf b*} we add a
prime to the symmetry labels for the b-axis branches). They are
sometimes simply labelled as longitudinal optic branches. However,
this labelling is rather imprecise as there are other phonon
branches with a different symmetry having a longitudinal character
as well. On the other hand, the polarization may be very different
for different branches of the same symmetry class: as will be
discussed below, some of the $\Delta_1$($\Delta_1$')branches are
c-axis polarized and hence are not observable for momentum
transfers within the [100]-[010]-scattering plane. Nevertheless,
they cannot be ignored in the interpretation of our data because
they hybridize with in-plane polarized branches in certain parts
of the Brillouin zone (because they belong to the same symmetry
class) and hence show up in the scans under discussion.

Our conclusions regarding c-axis-polarized $\Delta_1$ branches are
based on previous studies. In particular, the c-axis polarized
branches with zone-center energies of 62 meV and 54 meV,
respectively, were studied in Ref. 14, where they were measured
with a large c*-component of the momentum transfer. A summary of
the available data from the present and from previous
studies\cite{Pini1998} together with the calculated dispersion
curves in the whole range of phonon energies is shown in Fig. 6.

As already indicated in Sec. II, a very important step in
establishing the dispersions of the bond-stretching modes has come
from studies of branches with $\Delta_4$-symmetry, i.e. the modes
where the atoms in the Cu-O bi-layer move in opposite phase. We
found that the bond-stretching vibrations of $\Delta_1$-symmetry
and those of $\Delta_4$-symmetry have very similar energies, as
one should expect from the weak coupling across the Y atom. What
makes a study of the $\Delta_4$-bond-stretching branches very
attractive is the fact that mixing with c-axis polarized modes is
less of a problem. The $\Delta_4$-modes had to be studied on a
twinned sample. The measurements were made around the (3,0,2)
reciprocal lattice point. Representative spectra are shown in Fig.
7; see Ref. 17 for further data. As one can see, high resolution
allowed us to separate the modes for the {\bf a*} direction and
for the {\bf b*} direction, respectively. The results for the {\bf
a*} direction are summarized in Fig. 8. The results for the {\bf
b*} direction are included in Fig. 1.

The good agreement between our model and the experimental gives
support to a rather conventional picture of phonon dispersion with
a notable exception of anomalous dispersion of plane oxygen
bond-stretching vibrations calling for the special terms. The
following picture emerges from our analysis: the branches starting
at E=67 meV and 73 meV, respectively, have in-plane Cu-O
bond-stretching character with the atoms in the Cu-O bi-layer
moving in-phase. The energy difference at the zone center can be
explained entirely by the difference in the Cu-O bond lengths
along a and b. Both branches show a steep downward dispersion
half-way to the zone boundary. The b-polarized branch even has a
pronounced local minimum at q$\approx$0.27 r.l.u. whereas the
a-polarized branch does not have such a local minimum but is quite
flat for q$>$0.3 r.l.u. As will be discussed in a later section,
the local frequency minimum in the b-polarized branch is seen only
at low temperatures. The dispersion of the bond-stretching modes
with $\Delta_1$-symmetry is, however, somewhat obscured by an
anticrossing with a c-axis polarized branch which starts from the
Ag mode at 62 meV. In the {\bf b*} direction, a further
complication arises from the presence of another branch of the
same symmetry in the energy region of interest: the corresponding
zone-center mode has an energy of E=59 meV and involves
longitudinal Cu-O chain bond-stretching vibrations. When
approaching the zone boundary, there is another problem for the
determination of the bond-stretching phonon frequencies: the
intensities associated with bond-stretching modes overlap with
those from other branches dispersing upwards in the energy range
45-55 meV (Figs. 4,5). They have, in principle, in-plane oxygen
bond-bending character, but there will be inevitably some
hybridization with the bond-stretching mode once the two modes are
very close in energy. We note that this hybridization vanishes
only close to the zone boundary, because there the bond-bending
modes and the bond-stretching modes belong to different symmetry
classes.

Motivated by reports\cite{Chung} of an unusual dispersion of
transverse bond-stretching branches in YBa$_2$Cu$_3$O$_7$ we
studied these branches as well. As can be seen in Fig. 9, our data
do not support the claim that the transverse branch along {\bf b*}
has a rather strong dispersion up to q=0.25; to the contrary, this
branch has as little dispersion as the corresponding one along
{\bf a*}\cite{difference}. Figure 9 also shows the TO branch
starting from the longitudinal chain O mode. There was no problem
to follow this branch up to the zone boundary but attempts to
follow the corresponding branch in the longitudinal direction had
only very limited success. We learned from our model calculations
that the LO branch starting from the chain O mode hybridizes very
strongly with the other branches of the same symmetry - in
particular, with the c polarized apical O vibrations - already at
small wave vectors. For this reason, we were unable to follow this
branch beyond q=0.15 r.l.u.

\subsection{Temperature dependence}
Selected phonons were studied as a function of temperature on a
twinned sample. We found that the frequencies of the zone center
phonons and of the zone boundary phonons show very little change
when going from T=12 K to room temperature (Fig. 10). A certain
increase in linewidth as is evident for the zone boundary phonon
is typical of anharmonic behavior. The decrease of the phonon
intensities with increasing temperature can be accounted for by
the Debye-Waller-factor. Our finding that even the strongly
renormalized zone boundary phonons show very little temperature
dependence was in conflict with claims made by Chung et
al.\cite{Chung} that the bond-stretching phonons show an anomalous
temperature behavior throughout the Brillouin zone (see Fig. 9 of
Ref. 15). Therefore, we finally embarked on a detailed study to
check these claims. Although we could not confirm many of the
temperature induced changes reported in Ref. 15 we found indeed a
very pronounced effect of certain phonon modes. As explained in a
separate publication\cite{Pinicondmat} a downward shift of
spectral weight by at least 10 meV was observed for b polarized
bond-stretching phonons of $\Delta_4$'-symmetry at {\bf
q}$\approx$0.27 r.l.u. on cooling from room temperature to T=12 K.
The temperature evolution starts well above the superconducting
transition temperature, i.e. at about T=200 K. Data for T=200 K
have been included in Fig. 1. These data were important for
adjusting the parameters of our model. When the parameters were
adjusted to better fit the data of the $\Delta_4$'-branches, we
obtained an improved fit for the $\Delta_1$'-branches as well,
which further validates our analysis.

\section{Discussion}
The present results qualitatively confirm the results of an early
study by Reichardt\cite{Reichardt1996} but also go much beyond it
in that we were able to elucidate the a-b anisotropy of the
bond-stretching modes. Furthermore, we have gained some insight
into the origin of the broad intensity distributions observed in
the region half-way to the zone boundary. It appears that that the
complex lineshapes observed for a-polarized bond-stretching modes
of $\Delta_1$-symmetry can be attributed in large part to
anticrossings with other branches of the same symmetry. We agree
with the observation reported in Ref. 15 that the bond-stretching
branch along {\bf a*} splits into two branches half-way to the
zone boundary; however, we disagree on the interpretation. The
phenomenon is the natural consequence of hybridization between the
O in-plane bond-stretching modes and those of apical O vibrations
along c. This phenomenon is illustrated in Fig. 11 for the zone
boundary point. As for the b direction, there are two reasons for
the observed complex line shapes: (i) anticrossings of different
branches and (ii) the anomalously steep dispersion - probably
associated with very broad intrinsic linewidths - at {\em q}
half-way to the zone boundary. In order to study the importance of
(ii) we simulated the scans shown in Figs. 3,5 using the model
fitted to the 200 K data, i.e. a model giving a smooth dispersion
throughout the Brillouin zone. The following conclusions are drawn
from this simulation (Fig. 12) Although the differences between
Fig. 12 and Fig. 5b are somewhat subtle, there is definitely a
better agreement between the experimental results displayed in
Fig. 5a and the calculated ones in Fig. 5b than with those in Fig.
12. This indicates that the bond-stretching branch of
$\Delta_1$'-symmetry exhibits the same type of anomalous low-T
behaviour as its counterpart of $\Delta_4$'-symmetry. This
conclusion is in line with evidence for an anomalous temperature
dependence of $\Delta_1$'-modes reported in Ref. 15. On the other
hand, it would have been extremely difficult to establish details
of the temperature dependence from measurements of the
$\Delta_1$'-branches alone in view of the formidable complication
due to the anticrossings. Twinning of the sample further
aggravates the situation considerably. Therefore, it is not
surprising that the temperature dependence reported in Ref. 15
agrees with our results only on a qualitative level.

Measurements of the O in-plane bond-stretching phonons in
optimally doped La$_{1.85}$Sr$_{0.15}$CuO$_4$ revealed a
considerable broadening of the phonon lines for q$>$0.1
r.l.u..\cite{Pini1998,Pini1999}. This broadening was found to be
particularly pronounced around q=(0.3,0,0)\cite{Pini1999}. We
expect a similar effect for optimally doped YBCO. Unfortunately,
the problems for extracting linewidths discussed above do not
allow us to deduce intrinsic phonon linewidths from the available
data. We hope to achieve progress on this matter by a study of the
$\Delta_4$ resp. the $\Delta_4$' modes on a detwinned sample.

Our model calculations indicate that the observed a-b anisotropy
of the bond-stretching branches can be largely attributed to the
orthorhombicity of the structure (apart from the anomalous
temperature dependence along {\bf b*})\cite{agree}. That is to
say, the general energy difference between the two directions is
reproduced by a simple potential model in conjunction with the
experimental difference between the a and b lattice parameters.
Further, the special terms needed to lower the zone boundary
frequencies are of similar size. This contrasts with the results
of a previous study of underdoped YBa$_2$Cu$_3$O$_{6.6}$ which had
shown a much stronger {\bf a-b} anisotropy at the zone
boundary\cite{Pini2002} which remains to be understood.

We have already mentioned in Sec. III the results of a recent
ab-initio calculation \cite{Bohnen} of the phonon dispersion in
YBa$_2$Cu$_3$O$_7$ using Density Functional Theory in the local
density approximation. The theory quantitatively predicts the
frequency drop from the zone center to the zone boundary for the
bond-stretching modes both along {\bf a*} and {\bf b*} at T=200 K;
however, it also predicts a considerable downward dispersion for
the [110] direction as well ($\approx$8 meV) which is not borne
out by experiment\cite{Reichardt1996}. (Also, it is not possible
to check the doping dependence because of the known limitations of
LDA for describing the correlated-insulator parent compound,
YBa$_2$Cu$_3$O$_6$.) Turning to a different mode, the theory makes
an accurate prediction of a rather low frequency for the
longitudinal chain O vibrations (61.5 meV). This is not quite as
low as observed in experiment\cite{note} (59 meV) but much lower
than expected from the simple potential model\cite{Chaplot} (75
meV) in which the Cu-O potential was assumed to be the same for
the Cu-O chains and for the Cu-O planes. Unfortunately, we were
unable to check the theoretical predictions that the longitudinal
branch starting from the chain O mode has a strong downward
dispersion and exhibits a local minimum at about half-way to the
zone boundary, related to Fermi surface nesting of chain O
electronic states. The strong mixing of the chain O mode with
other modes of the same symmetry did not allow us to isolate the
contributions of the chain O vibrations from the rest.

We note that in another cuprate, i.e. (La,Sr)$_2$CuO$_4$, the
phonon renormalization upon doping is also much stronger in the
[100] direction than in the [110] direction\cite{Pini1998}. Thus,
this behavior seems to be common to Cu-O planes, but independent
of extraneous features such as chains. The doping dependence and
spatial anisotropy of the phonon softening appear to be captured
by a calculation in which the t-J model is extended to explicitly
include electron-phonon couplings.\cite{Roesch,Horsch}. R\"osch
and Gunnarsson\cite{Roesch} have evaluated the doping and q
dependence of the O bond-stretching mode of a Cu-O layer through a
t-J model with electron-phonon interactions derived from a
three-band model. With their purely two-dimensional model, they
obtain the correct anisotropy between the [1,1] and [1,0]
directions, and reasonable results for the doping- and q-dependent
softening of the mode along the [1,0] direction.

From the discussion above, it is evident that the lattice dynamics
of YBa$_2$Cu$_3$O$_7$ is less anomalous than advocated in Ref. 15,
but nevertheless we think that the behavior of the bond-stretching
modes indicates a strong electron-phonon coupling. The question of
whether or not this electron-phonon coupling is relevant for high
T$_c$ superconductivity cannot be directly answered from the
phonon data alone. From their theoretical analysis, Bohnen, Heid,
and Krause\cite{Bohnen} came to the conclusion that conventional
electron-phonon coupling is much too weak to explain a high T$_c$,
in contrast to the case of MgB$_2$ where conventional
electron-phonon does seem to explain the observed
T$_c$\cite{Bohnen2001}.

\section{Conclusions}
The present study has elucidated the {\em a-b} anisotropy of the
plane-polarized bond-stretching vibrations in optimally doped
YBa$_2$Cu$_3$O$_{6.95}$. We confirmed that the bond-stretching
branches show a steep downward dispersion in both the a and the b
directions at q$\approx$0.25 r.l.u. that leads to anti-crossings
with {\em c} polarized branches. The resulting anti-crossing gaps
can be explained by lattice dynamical calculations using a shell
model. In the a direction, the bond-stretching mode of
$\Delta_1$-symmetry is strongly hybridized with a c-polarized
branch over a large part of the Brillouin zone, which explains the
''splitting'' of this branch reported in Ref. 15 in a natural way.
The dispersion of the bond-stretching branches is very well
reproduced by density-functional theory\cite{Bohnen} if the
comparison is based on 200 K data, whereas the pronounced
softening observed in b-polarized branches at low T remains
unexplained. Although the pronounced renormalization of zone
boundary bond-stretching modes on going from insulating O6 to
superconducting O7 indicates a strong electron-phonon coupling,
theory\cite{Bohnen} suggests that it is too weak to be relevant
for high T$_c$ superconductivity. Thus, the temperature effect,
presented in detail in Ref. 17, seems to be the most anomalous
phenomenon in the phonon properties of optimally doped YBCO.

\begin{acknowledgments}
We are indebted to Dr. Y. Shiohara and to Dr. S. Koyama at
Superconductivity Research Laboratory for their help in growing
the YBCO crystals. This work was partially supported by the New
Energy and Industrial Technology Development Organization (NEDO)
as Collaborative Research and Development of Fundamental
Technologies for Superconductivity Applications. JMT is supported
by the U.S. Department of Energy's Office of Science under
Contract No. DE-AC02-98CH10886. This work was partially supported
by the Japan Science and Technology Agency within the CREST
project. This project has been also supported by the Grant in Aid
for Scientific Research from the Japan Society of Promotion of
Science.
\end{acknowledgments}

\appendix
\section{Appendix A: Anti-Crossing Behavior}
YBa$_2$Cu$_3$O$_7$ has 13 atoms in the unit cell and hence there
are 39 branches of the phonon dispersion in each direction. These
branches can be grouped according to symmetry of the eigenvectors
of the atomic displacements. For the case of the [100] or the
[010] direction, there are four different symmetry classes. Within
each subset of dispersion curves, the polarization may be very
different. Still, none of the branches is allowed to cross another
one of the same subset. If, as a result of the special nature of
the interatomic interactions, two branches tend to cross each
other in a particular part of the Brillouin zone, this will lead
to a mixing of the polarization patterns and to a mutual repulsion
of the eigenfrequencies (exactly as in the problem of coupled
harmonic oscillators). However, the degree of hybridization and
the strength of the repulsion cannot be predicted from general
principles but depend on the symmetry of the crystal lattice and
the peculiarities of the interatomic interactions. For
illustrative purposes, we consider the following case: a simple
interatomic potential model\cite{Chaplot} predicts a flat
dispersion for the two topmost branches of $\Delta_1$-symmetry in
the {\em a} direction (Fig. 13). These two branches have in-plane
polarization and c-axis polarization, respectively. As a
consequence, the inelastic scattering structure factors calculated
for momentum transfers along the line (3+x,0,0), i.e. in the basal
plane, will be very different for the two branches. In a
corresponding neutron scattering experiment, only the upper branch
will be detected. The situation dealt with so far corresponds to
the situation found in O6. In O7, however, the upper branch
acquires strong dispersion from electron-phonon coupling effects
as discussed in this paper. For the sake of simplicity, we have
simulated such effects by adding a single special term to the
dynamical matrix lowering the frequency of only the
bond-stretching vibrations.The term is designed to produce a
maximum effect at q=0.25 r.l.u. (for details, see APPENDIX B).
This term was tuned to reduce the frequency of the bond-stretching
vibrations at q=0.25 to below that of the apical O vibrations. As
a consequence, phonons of the lower branch acquire some {\em
a}-axis polarization over an extended range of wave vectors making
them observable at momentum transfer along a (Figure 13).
Evidently, hybridization and repulsion of the two types of
vibrations are quite strong in YBa$_2$Cu$_3$O$_7$.

\section{Appendix B: Special Force Constants}
As mentioned in Section III.1, several special terms have to be
added to the shell model to arrive at a quantitative description
of the bond-stetching branches. The term describing screening by
free carriers has been explained in detail in Ref. 30. The term
accounting for a planar breathing deformability has been given in
Ref. 30 as well. The parameter A for this planar breathing
deformability was tuned to reproduce the
observed\cite{Reichardt1996} bond-stretching phonon dispersion in
the [110] direction. Although the bond-stretching modes in the
[110] direction do show a considerable renormalization with the
metal-insulator transition, the branch remains flat. As a
consequence, the parameter A was found to be rather small (A=25000
dyn/cm).

The further terms needed to arrive at a satisfactory description
of the bond-stretching branches in the [100] and in the [010]
directions are {\em ad hoc} extensions of the term for the
breathing deformability: a 2x2 sub-matrix is added to the
dynamical matrix for the plane oxygen atoms O2 and O3 (and, of
course, for the other plane oxygen atoms in the bi-layer as well).
The diagonal elements of the matrix are given by
\begin{displaymath}
D_2(1,1) = B_{100} \times sin^2 (\pi q_x)+ C_{100}\cdot sin^2
(2\pi q_x)) \cdot|sin^2 (\pi q_x)-sin^2(\pi q_y)|^{1/2}
\end{displaymath}

\begin{displaymath}
D_2(2,2) = B_{010}\times sin^2 (\pi q_y)+C_{010}\cdot sin^2(2\pi
q_x)) \cdot|sin^2(\pi q_y)-sin^2(\pi q_y)|^{1/2}
\end{displaymath}

The non-diagonal elements of the matrix are zero. For negative
values of B$_{100}$ the frequencies of the bond-stretching modes
are lowered in the [100] direction in a sine-like manner from the
zone center to the zone boundary. On the other hand, this term has
no effect in both the [010] and in the [110] directions. From a
fit to the experimental data we obtained B$_{100}$=-70000 dyn/cm.
The parameter C$_{100}$ describes the second harmonic of the
linear breathing deformability with a maximum effect at {\bf
q}=0.25 {\bf a*} and zero effect at {\bf q}=0 and at {\bf q}=0.5
{\bf a*}. The fit resulted in a value C$_{100}$=-60000 dyn/cm. The
corresponding terms for the [010] direction were obtained as
B$_{010}$=-75000 dyn/cm and C$_{010}$=-55000 dyn/cm.

In order to describe the anomalous softening upon cooling in the
[010] direction at least in a semi-quantitative manner another
term was added to D$_2$(2,2) in analogy to the term with the
prefactor C$_{010}$ but with the factor sin$^2$(2$\pi$ qy) raised
to the nth power. This term was adjusted by trial and error to
C'$_{010}$=-100000 dyn/cm and n=6.

\newpage

\clearpage
\begin{figure}
\caption{Illustration of the effect of the special terms included
in the model calculations. Black line: no extra terms. Green line:
after inclusion of a (negative) planar breathing deformability
chosen such as to reproduce the experimental zone boundary
frequency in the (110) direction. Violet line: after inclusion of
a special term to lower the frequency of the 'half-breathing'
mode. Red line: after inclusion of a special term to lower the
frequency of the bond-stretching modes halfway to the zone
boundary. Blue line: after inclusion of a further term to simulate
the anomalous softening at low temperatures. Only the branch with
in-plane bond-stretching character is shown for clarity. Note that
the special terms directly affect only the vibrations with
bond-stretching character. Blue/Red circles denote phonon peak
positions at T=12K/200K. The blue bar at q=0.275 r.l.u. denotes an
ill-defined phonon peak containing also contributions from the
next lower phonon branch with bond-bending character.}
\label{firstfigure}
\end{figure}

\begin{figure}
\caption{Measured (a) and calculated (b) neutron scattering
intensity along the a direction from Q=(4,1,0) to (3.5,1,0). This
Brillouin zone maximizes the scattering cross section for phonons
with strong longitudinal plane oxygen vibrations perpendicular to
the chains. Further, it allows one to exploit focusing effects for
branches with a downward dispersion. Background, linear in energy
and independent of Q, has been subtracted from the experimental
raw data to emphasize one-phonon scattering. The calculated
one-phonon neutron scattering intensity was corrected for
spectrometer resolution.}
\label{secondfigure}
\end{figure}

\begin{figure}
\caption{The same as Fig.1 but for the b direction. The
measurements were carried out from Q=(1,4,0) to (1,3.5,0).}
\label{thirdfigure}
\end{figure}

\begin{figure}
\caption{Coulour coded contour plots of the measured and
calculated neutron scattering intensities shown in Fig. 1.  No
data is available for the regions colored in brown. The calculated
phonon dispersion curves (without the resolution correction) are
shown as blue lines. According to the model, phonon branch
anticrossing causes dramatic changes in eigenvectors within the
same branches between the zone center and the zone boundary, which
results in corresponding changes in one-phonon scattering cross
sections. Apart from this effect, there is a 30 \% increase of the
structure factors of the bond-stretching modes from the zone
center to the zone boundary in the (4,1,0) Brillouin zone.}
\label{fourthfigure}
\end{figure}

\begin{figure}
\caption{The same as Fig. 3 but for the b direction. The
measurements were carried out from Q=(1,4,0) to (1,3.5,0).}
\label{fifthfigure}
\end{figure}

\begin{figure}
\caption{Calculated dispersion curves of phonon branches of
$\Delta_1$-symmetry and of $\Delta_1$'-symmetry, respectively, as
well as experimentally measured phonon peak energies. Arrows point
to Ag modes, the other modes being of B$_{3u}$ (B$_{2u}$)
character. Data in the energy range 10 meV$\leq$E$\leq$42 meV were
taken from previous measurements on twinned
samples\cite{Pini1998}. Data for the longitudinal acoustic
branches were taken from previous measurements on a small (12
mm$^3$) detwinned sample\cite{Pyka}.} \label{sixthfigure}
\end{figure}

\begin{figure}
\caption{Neutron spectra taken on a twinned sample at T=12 K with
a high resolution configuration. The spectra were fitted with two
resp. three Gaussians and a sloping background. The peaks are
assigned to branches of $\Delta_4$-symmetry in the {\em a}
direction or the {\em b} direction, respectively.}
\label{seventhfigure}
\end{figure}

\begin{figure}
\caption{Dispersion of high-energy phonon branches of
$\Delta_4$-symmetry along the a direction as determined on the
twinned sample. The full lines were calculated from the model
described in the text and the dashed lines were taken from Ref.
23. Note that the {\em ab-initio results}\cite{Bohnen} were
shifted upwards by 1 meV.} \label{eigthfigure}
\end{figure}

\begin{figure}
\caption{Dispersion of the transverse bond-stretching branches
along {\bf a*} and along {\bf b*}, respectively, as observed on an
untwinned sample at T=12 K. The high-energy branch and the low
energy branch along {\bf a*} have in-plane oxygen and chain oxygen
character, respectively. Lines are a guide to the eye. The arrows
denote the frequencies observed by ellipsometry\cite{Bernhard}.}
\label{ninthfigure}
\end{figure}

\begin{figure}
\caption{Energy scans taken on a twinned sample at two different
temperatures.}
\label{tenthfigure}
\end{figure}

\begin{figure}
\caption{Displacement patterns of the two highest energy zone
boundary modes with wave vector q=(0.5,0,0) as calculated from two
different models A and B. Model A is described in the text and
reproduces the anomalously low frequency of the in-plane Cu-O
bond-stretching mode by including special terms designed for that
effect. This choice leads to a strong hybridization of the
in-plane bond-stretching mode and of the c-axis polarized apical O
mode with energies of 58 meV and 66 meV, respectively. It
corresponds to the situation displayed in Fig. 5. The special
terms are ommitted in model B which leads to a much higher
bond-stretching mode frequency (74 meV) and consequently to a much
smaller hybridization with the apical O mode (61 meV).}
\label{eleventhfigure}
\end{figure}

\begin{figure}
\caption{Colour coded contour plot of calculated 200K neutron
scattering intensity for the line from Q=(1,4,0) to (1,3.5,0). The
calculated phonon dispersion curves along the high symmetry line
are shown as blue lines. The model used for this plot differs from
the one used for Fig. 5 by omitting the special term introduced to
describe the anomalous softening observed at low temperatures. The
corresponding behavior of the $\Delta_4$'-branches is shown in
Fig. 1 by a red line and a blue line, respectively.}
\label{twelfthfigure}
\end{figure}

\begin{figure}
\caption{Top: Calculated dispersion of the two highest
$\Delta_1$-branches before (full lines) and after (dashed lines)
inclusion of a special term to reduce the frequencies of the bond-
stretching modes half way to the zone boundary. Bottom: inelastic
structure factors calculated along the line (3,0,0) to (3.5,0,0).}
\label{thirteenthfigure}
\end{figure}

\begin{table}\caption{Table I: Parameters of the shell model for YBa$_2$Cu$_3$O$_7$.
Z(k), ionic charge;
Y(k), shell charge; K(k), core shell force constant. The repulsive
Born-Mayer potential used in Ref. 19 was replaced by force
constants: F$_{kk'}$ and G$_{kk'}$ are longitudinal and transverse
force constants, respectively. Following Ref. 19, an attractive
van der Waals potential with C$_{kk'}$=100 eV \AA was assumed to
act between oxygen atoms. Note that the chain copper and the chain
oxygen are labelled as Cu(1) and O(4), respectively.}

\begin{ruledtabular}
\begin{tabular}{cccc}
k & Z(k) & Y(k) & K(k)(nm$^{-1}$)\\
\hline\hline 1=Y & 1.85 & 7.8 & 8000\\
2=Ba & 1.87 & 5.6 & 1200\\
3=Ca(1) & 1.52 & 2.8 & 2000\\
4=Cu(2,3) & 1.83 & 4.2 & 2000\\
5=O(1) & -1.62 & -2.8 & 1200\\
6=O(2,3) & -1.565 & -2.2 & 1200\\
7=O(4) & -1.26 & -2.2 & 1200\\
\hline\\
r(\AA) & k,k' & F$_{kk'}$(dyn/cm) & G$_{kk'}$(dyn/cm)\\
\hline\hline\\
1.859 & 3,6 & 375000 & -39000\\
1.9286 & 4,5 & 294626 & -33892\\
1.9418 & 3,7 & 202706 & -21184\\
1.9611 & 4,5 & 264029 & -30533\\
2.2845 & 4,6 & 81514 & -4615\\
2.3810 & 1,5 & 95669 & -13177\\
2.4075 & 1,5 & 95518 & -8114\\
2.6882 & 6,7 & -23182 & 0\\
2.7229 & 5,5 & 10855 & -919\\
2.7380 & 2,6 & 77069 & -8662\\
2.8464 & 5,5 & 9734 & 0\\
2.8720 & 2,7 & 53902 & -3158\\
2.9678 & 2,5 & 57065 & -8451\\
2.9890 & 2,5 & 44574 & -5715\\
3.1927 & 5,6 & 2500 & 0\\
3.2090 & 1,4 & 0 & 0\\
3.2125 & 5,6 & 2500 & 0
\end{tabular}
\end{ruledtabular}
\end{table}


\begin{thebibliography}{10}
\bibitem{difference}We note that there is a systematic difference between the
data of ref. 15 and our data of about 1 meV which we attribute to
a slightly different calibration of the neutron spectrometers. Our
calibration is supported by the good agreement with the optical
data of ref. .
\bibitem{note}We note that the tentative assignment of weak peaks observed in
Ref.  at energies $\approx$70 meV and $\approx$61 meV,
respectively, to the chain O modes and to c-axis displacements of
apical O has to be interchanged.
\end{thebibliography}
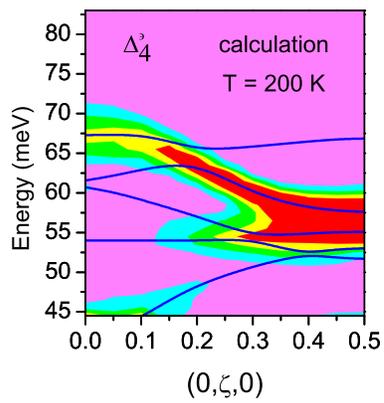
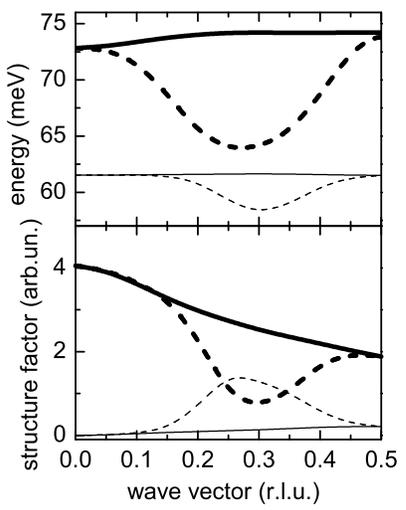
\end{document}